# Unveiling the detection dynamics of semiconductor nanowire photodetectors by terahertz near-field nanoscopy

Eva A.A. Pogna,[1] Mahdi Asgari,[1] Valentina Zannier,[1] Lucia Sorba,[1] Leonardo Viti,[1] and Miriam S. Vitiello[1]

[1] *NEST, CNR - Istituto Nanoscienze and Scuola Normale Superiore, Piazza San Silvestro 12, 56127, Pisa, Italy*

**Semiconductor nanowire field-effect transistors represent a promising platform for the development of room-temperature (RT) terahertz (THz) frequency light detectors due to the strong nonlinearity of their transfer characteristics and their remarkable combination of low noise-equivalent powers (< 1 nW/Hz$^{1/2}$) and high responsivities (> 100 V/W). Nano-engineering a NW photodetector combining high sensitivity with high speed (sub-ns) in the THz regime at RT is highly desirable for many frontier applications in quantum optics and nanophotonics, but this requires a clear understanding of the origin of the photo-response. Conventional electrical and optical measurements, however, cannot unambiguously determine the dominant detection mechanism due to inherent device asymmetry that allows different processes to be simultaneously activated. Here, we innovatively capture snapshots of the photo-response of individual InAs nanowires via high spatial resolution (35 nm) THz photocurrent nanoscopy. By coupling a THz quantum cascade laser to scattering-type scanning near-field optical microscopy (s-SNOM) and monitoring both electrical and optical readouts, we simultaneously measure transport and scattering properties. The spatially resolved electric response provides unambiguous signatures of photo-thermoelectric or bolometric currents whose interplay is discussed as a function of photon density and material doping, therefore providing a route to engineer photo-responses by design.**

## Introduction

Photo-detection relies on the ability to convert absorbed photons into a stable electrical signal[1]. Photocurrent generation at terahertz (THz) frequencies in semiconducting nanostructures can be mediated by physical mechanisms such as the bolometric effect,[2–4] photo-thermoelectric effect (PTE),[1,5,6] photovoltaic effect[7], galvanic effect[8,9] and excitation of plasma waves (PWs)[10–13].

InAs nanowires (NWs) have proven to be an efficient active material for the development of state-of-the-art field-effect transistors (FETs) operating as room-temperature (RT) THz photodetectors[14,15] due to their high electron mobility[16,17] (up to 6000 cm$^2$V$^{-1}$s$^{-1}$), attofarad (aF) capacitance[16,17], and high saturation velocity[18] (~1.3 10$^7$ cms$^{-1}$ at a field of 16 kVcm$^{-1}$), which can easily enable high transconductance at low gate voltages and high cut-off frequencies[18] (~1 GHz). Furthermore, their narrow bandgap and degenerate Fermi-level pinning allow for the easy formation of stable ohmic contacts, which become increasingly important as the transistor is scaled down.

While in the last few years, the dependences of the THz detection performance of semiconductor NW photodetectors on material-related parameters (NW doping[19], carrier concentration, geometry, and/or material choice) have been widely investigated[20], a systematic study of the physical mechanisms at the base of RT photodetection is still lacking.

Present assumptions indeed only rely on what can be indirectly inferred by mapping the gate-tunable conductivity through transport experiments, analyses of photo-responses in antenna-integrated devices[21] or on-chip photocurrent pump-probe studies[22].

Mapping the spatial distribution of the photocurrent in an active material conversely represents an unambiguous way to discriminate the dominant process controlling photo-detection since individual physical mechanisms correspond to distinctive profiles of the light-induced photocurrent.
Scanning photocurrent microscopy based on far-field optics has been successfully applied in the last few years to investigate local photocurrents induced by visible and near/mid-infrared illumination in one-dimensional (1D) nanostructures such as NWs[13,23] and nanotubes[24–26].

The extension of photocurrent microscopy to the THz frequency range, an interesting frontier for the rapidly emerging development of micro- and nanosources[22,27], nanodetectors[14,20,25,28], modulators[29–31] and metamaterials[32], has been, however, traditionally hindered by the diffraction limit. Overcoming this limitation, near-field techniques can have a tremendous impact on engineering the aforementioned technologies with progressively improved performances and functionalities.

Near-field optical microscopy has shown amazing potential in the last few years for investigating optoelectronic properties of nanoscale materials and devices thanks to its unique capabilities of inspecting charge carrier density[13,33–37] and plasmon-polariton[38–41] and phonon-polariton[42–44] modes[45] with unprecedented spatial resolution. Furthermore, near-field microscopy can allow mapping of the spatial variation and the bias dependence of local currents induced by light illumination, therefore tracking the photocarrier transport and electronic band bending in electronic and photonic nanodevices.

Very recently, scattering-type scanning near-field optical microscopy (s-SNOM) has been exploited at THz frequencies to map photocurrents in graphene[46] with sub-diffraction spatial resolution, exploiting lightning-rod effects in a metal-coated atomic force microscopy (AFM) tip illuminated by the THz output of a bulky, table-top, gas laser[47].

Here, we conceive and devise a compact, portable, scanning near-field system that functions at THz frequencies and is capable of simultaneously performing photocurrent nanoscopy[48-59] and detectorless near-field imaging[60] of a 1D NW field-effect nanodetector with ~35 nm spatial resolution, which corresponds to a wavelength fraction < $\lambda/3000$. The system comprises a continuous-wave (CW) THz quantum cascade laser (QCL) coupled to s-SNOM and is here innovatively exploited to trace and map, unambiguously, the physical mechanisms inducing light detection at THz frequencies.

**Results**

Our samples employ Au-catalysed InAs NWs slightly doped with Se grown by chemical beam epitaxy following a *bottom-up* approach[61]. The resulting crystalline phase is wurtzite, and the InAs NWs exhibit {110} facets and hexagonal cross-sections with lengths ⸰2.5 µm and radii ≈ 40 nm. Following the fabrication procedure described in the Methods section, the individual NWs are integrated in lateral-gated FETs.

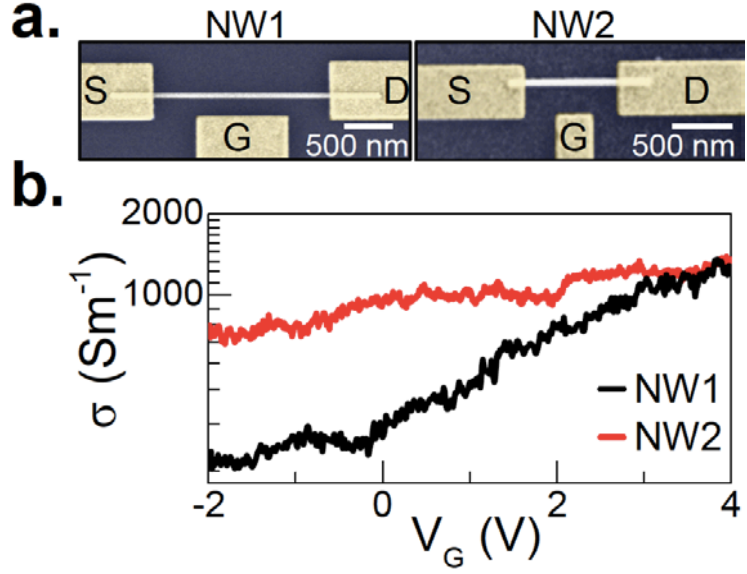

**Figure 1: (a)** Scanning electron microscopy (SEM) images of InAs NW field effect transistors labelled NW1 and NW2, with source (S), drain (D) and gate (G) contacts indicated. **(b)** The conductivity σ of NW1 (black line) and NW2 (red line) measured at room temperature applying drains to source voltage of $V_{DS}$= 2 mV for NW1 and $V_{DS}$= 5 mV for NW2, reported as a function of the gate voltage $V_G$.

Figure 1a shows scanning electron microscopy (SEM) images of two prototypical devices, each consisting of a single NW-FET with source (S), drain (D) and gate (G) contacts as indicated. The two devices, denoted NW1 and NW2, differ in the gate-NW distance $d$ ($d^{NW1}$ = 210 nm, $d^{NW2}$ = 267 nm) and the channel length $l$ ($l^{NW1}$ = 1.95 μm and $l^{NW2}$ = 0.86 μm). In both cases, the gate length $W$ ($W^{NW1}$=1 μm and $W^{NW2}$ = 328 nm) is smaller than the FET channel length. The FETs are primarily characterized through transport experiments, showing a stable ohmic behaviour, on-off current ratios higher than 10 (NW1) and 3 (NW2), and a lack of gate leakage through the SiO$_2$/Si substrate (see Supporting Information).

The conductivity $\sigma$ of the two devices as a function of the gate voltage $V_G$ is reported in Fig. 1b. The resistivities $\rho$ at $V_G$=0 V are $\rho_0^{NW1}$∼ 3 mΩ/m and $\rho_0^{NW2}$ ∼ 1.2 mΩ/m, and the transconductances $g$ in the linear regime are $g^{NW1}$ = 1.6 mS/m and $g^{NW2}$ = 13 mS/m. We estimate the capacitance $C$ between the lateral gate and the NWs for the two devices as $C^{NW1}$= 41 aF and $C^{NW2}$= 12 aF by numerical simulation based on a finite element method (FEM) using the *AC/DC* module of commercial software (Comsol Multiphysics). The carrier mobility $\mu$ in the two NWs can be extracted from the gate capacitance using the Wunnicke metrics[62] as $\mu = gl^2/(CV_{DS})$, giving $\mu^{NW1} = 740$ cm$^2$V$^{-1}$s$^{-1}$ and $\mu^{NW2} = 505$ cm$^2$V$^{-1}$s$^{-1}$ for the two devices. From the current dependence at the applied $V_{DS}$, we estimate the carrier density at zero gate voltage $n_e = (\mu e \rho_0)^{-1}$, where $e$ is the electron charge, obtaining the following estimations for the two devices: $n_e^{NW1} = 3\times10^{16}$ cm$^{-3}$ and $n_e^{NW2} = 1\times10^{17}$ cm$^{-3}$. The local modulation provided by the gate enables us to investigate the photo-response as a function of the carrier density with an expected carrier density modulation on the order of $10^{16}$ cm$^{-3}$V$^{-1}$.

The near-field photocurrent measurements are performed by combining scanning near-field THz nanoscopy with the electrical read-out at the S and D contacts of the FET, as sketched in Fig. 2a. The output of a THz-QCL emitting at 2.7 THz, operating in CW and linearly p-polarized, is focused on the 10 nm radius PtIr-coated AFM tip of an s-SNOM instrument, with a 30° incident angle relative to the sample surface. The AFM tip couples to the incident THz light, which is funnelled at its apex[63] and interacts with the approached sample. While performing a raster scan of the sample, we simultaneously monitor the topography with an infrared laser focused on the AFM probe cantilever and the D-S electrical signal. Specifically, we detect the D-S photovoltage $\varDelta$V when no $V_{DS}$ is applied ($V_{DS} = 0$ V) using a high input impedance (R=100 MΩ), low-noise voltage amplifier or, alternatively, the photocurrent $\varDelta$I for a finite $V_{DS}$ by using a low input impedance (R=50 Ω), low-noise (50 fA) transimpedance amplifier.

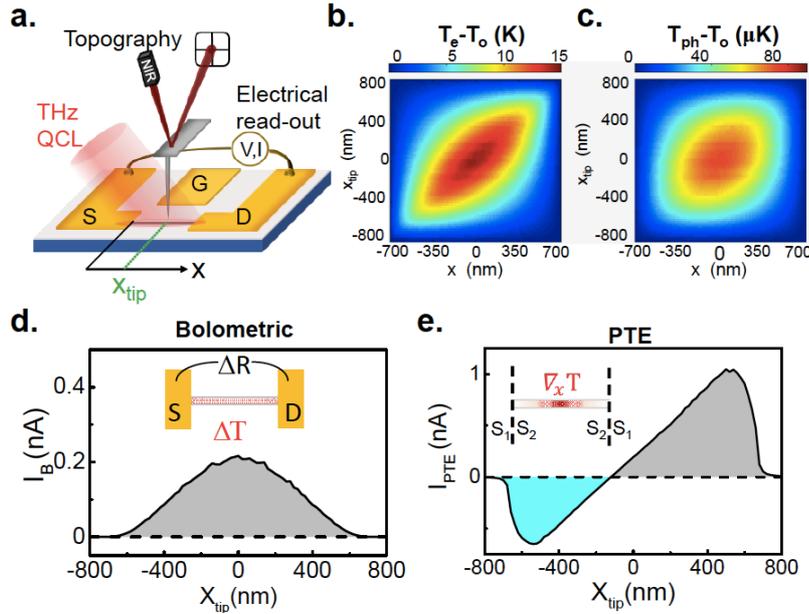

**Figure 2**: **(a)** Sketch of the photocurrent nanoscopy experiment. A THz-QCL coupled to the AFM tip of an s-SNOM photo-excites a single InAs NW integrated into a FET. Source (S), drain (D) and gate (G) contacts are marked on the graph. In the experiment, the S-D signal is monitored together with the sample topography. **(b-c)** Simulated local increases in the electron $T_e$ (b) and lattice $T_{ph}$ (c) temperatures with respect to the room temperature $T_o$ = 296 K induced by photo-excitation in a 1.4 μm long NW with a diameter of 80 nm and its centre at x = 0 as a function of the position $x$ along the NW axis and of the position $x_{tip}$ of the tip scanning the sample. **(d-e)** Simulated photocurrent profiles induced by the bolometric (d) and PTE (e) effects as a function of the tip position $x_{tip}$. The slightly positive shift of $x_{tip}$ at which the photocurrent value crosses zero reflects the near-field photo-excitation with the CW THz-QCL field funnelled by the AFM tip, which is illuminated from one side of the NW with a 30° incident angle.

Near-field photo-excitation with the CW THz-QCL field, impinging on the AFM tip with a 30° angle with respect to the sample plane, is funneled onto the NW, introducing an asymmetry along the NW axis and producing strongly localized heating in the NW caused by THz photon absorption via intraband transitions. Conversely, interband absorption can be discarded considering that the ☐356 meV energy gap of InAs[64] largely exceeds the THz photon energy (11.2 meV). The absorbed heat results in finite gradients of both $T_e$ and $T_{ph}$ along the NW axis that depend on the AFM tip position.

To predict the temperature profiles during a sample scan, we first simulate the electromagnetic field distribution along a NW via FEM simulation as a function of the tip position $x_{tip}$, and then we extract the power loss map (see Supporting Information). Interestingly, the simulated THz field power peaks below the tip apex and at the NW/electrode interfaces. The $T_e$ and $T_{ph}$ corresponding to a given $x_{tip}$, i.e., to a given distribution of the THz field, are then estimated by solving a 1D heat-diffusion model along the NW, including heat exchange with the substrate[65] (see Fig. 2b,c). Details on the thermal constants employed in the simulation are given in the Supporting Information. While the retrieved $T_e$ profiles along the NW axis peaked at $x_{tip}$ (see Fig. 2b), the $T_{ph}$, which describes the lattice temperature increase due to the coupling with the hot electron bath, presents a broader profile that is less dependent on $x_{tip}$ and peaked at the NW centre. The photocurrent signal at each $x_{tip}$ value is finally calculated from these temperature profiles, assuming thermally driven effects, i.e., bolometric and PTE photocurrents. The bolometric ($I_B$) and thermoelectric ($I_{PTE}$) photocurrents as a function of $x_{tip}$ are reported in Fig. 2d,e following the expressions:

$$I_B = \int_{-l/2}^{+l/2} \frac{d\sigma}{dT}(T_{ph} - T_0)\frac{dV}{dx}dx$$
$$I_{PTE} = \int_{-l/2}^{+l/2} \sigma S_e(T_e)\frac{dT_e}{dx}dx \qquad (1)$$

where $\sigma$ is the electrical conductivity, $S_e$ is the Seebeck coefficient and $x$ spans the NW length $l$. The two effects exhibit consistently different photocurrent profiles.

The expected near-field $I_B$ photocurrent profile (Fig. 2d) resembles the bell-shaped trend of $T_{ph}$, with a maximum near the NW centre and constant polarity. In contrast, the PTE voltage $V_{PTE}$ and the current $I_{PTE} = \sigma V_{PTE}$ exhibit a sign change along the NW, with zero signal near the NW centre. The bolometric current $I_B$ results from the variation of $\sigma$ with $T_{ph}$, as quantified by the bolometric coefficient $d\sigma/dT$. In contrast to PTE, $I_B$ could also arise when the device is homogeneously heated ($\nabla T = 0$) and depends only on the temperature variation from that under the dark condition ($T_o$ = 296 K). The polarity depends on the sign of the bolometric coefficient, which is negative (positive) for a degenerate (nondegenerate) semiconductor.

The PTE response instead emerges thanks to the combination of a finite $T_e$ gradient induced by the local photoexcitation and to a spatially varying Seebeck coefficient $S_e$. For the InAs NWs, a Seebeck coefficient $S_e^{NW} \sim 150$ μVK$^{-1}$ is extrapolated from the transfer characteristics (see Supporting Information), in good agreement with the recently determined temperature dependence of the transport properties of InAs NWs[66]. On the other hand, the Au electrodes have $S_e^{Au} \sim 0$ μVK$^{-1}$, so the interface electrodes/NW acts as a thermocouple, able to translate temperature gradients into an electrical signal. When the tip is at the centre of the NW, the two opposite PTE currents originating at the S and D electrodes cancel out, resulting in a clearly zero signal (Fig. 2e). The maximum PTE signal appears at a finite distance from the metal/NW interface since the difference ($T_e$-$T_o$) reaches its minimum on the metallic electrodes. The photovoltaic mechanism is not considered for InAs NWs because the THz incident photon energy is not compatible with interband transition photo-exciting electron-hole pairs. As expected, simulations confirm that the different photo-detection mechanisms that can come into play in InAs NWs correspond to clearly distinguishable near-field photocurrent profiles.

To perform photocurrent nanoscopy experiments, we operate our s-SNOM tip in tapping mode with an oscillating frequency Ω. To isolate the components of the electrical signals due to the near-field photo-excitation, *i.e.*, those induced solely by the light scattered by the tip, the signals are demodulated at high harmonics of the tapping frequency $Ω_n = nΩ$, *i.e.,* at the order *n* in the range *n* = 2−5. First, we collect near-field maps while setting $V_{DS}$= 0 V on sample NW1, with S grounded and collecting *ΔV* at the D electrode, as sketched in Fig. 3a. *ΔV* is the superposition of signals oscillating at the tapping frequency Ω and at its harmonics $ΔV_n$: $ΔV = \Sigma^5_{n=0} ΔV_n = ΔV_0 + ΔV_1 + ΔV_2 +…$; the second-order (*n* = 2) component $ΔV_2$ is shown in Fig. 3c for different values of $V_G$. A finite value is reported only when the tip scans the portion of the NW that is not covered by the metallic electrodes. Indeed, the THz near-field is screened by the metal contacts and fails to photo-excite the portion of NW buried beneath. The horizontal cut of each individual photovoltage map, taken by averaging over 50 nm (5 pixels) around the NW centre (as identified by the topographic image in Fig. 3a), is shown in Fig. 3e.

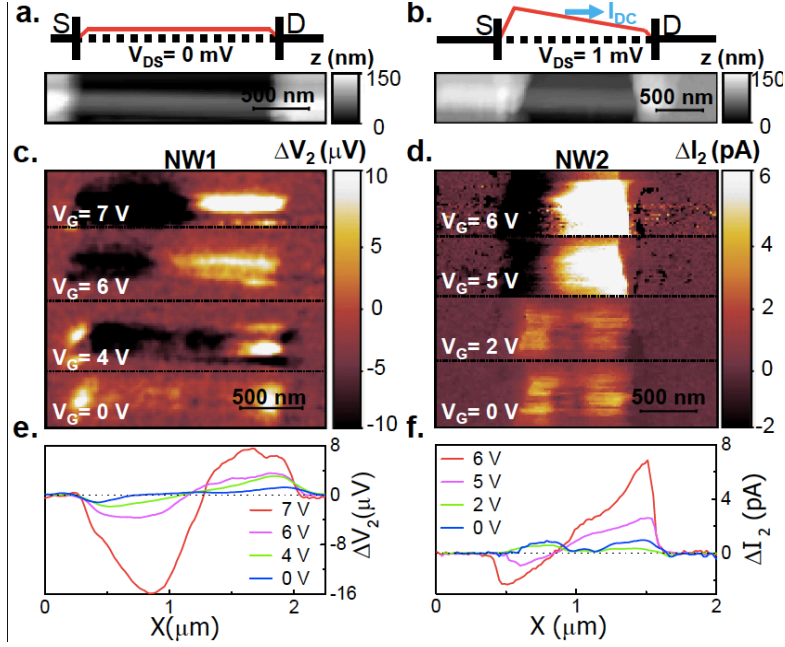

**Figure 3**: THz near-field photocurrent maps. **(a-b)** Sketches of the applied fields on the FET channel of NW1 with an applied $V_{DS} = 0$ V (a) and of NW2 with $V_{DS} = 1$ mV (b) together with the topographic maps (in greyscale) of the two FETs. **(c)** Near-field second-order photovoltage $\Delta V_2$ maps of NW1 with $V_{DS} = 0$ V and at different gate voltages in the range $V_G = 0\text{-}7$ V. **(d)** Near-field second-order photocurrent maps $\Delta I_2$ of NW2 with $V_{DS} = 1$ mV and $V_G$ in the range 0-6 V. **(e)** $\Delta V_2$ profiles of NW1 at different $V_{Gs}$ obtained by averaging the horizontal cuts of the maps in panel (c) over a 50 nm range around the NW1 centre. **(f)** $\Delta I_2$ profiles of NW2 at different $V_G$ values obtained by averaging horizontal cuts of the maps in panel (d) over a 50 nm range around the NW2 centre. The flat, nearly zero photocurrent region in panel e) reflects the gate-induced screening of the THz field at the centre of the NW.

In the maps and line profiles, the photovoltage exhibits a phase change along the NW; it increases in amplitude approaching the contacts and becomes negligible at the NW centre, as expected when photo-detection is governed by the PTE. Moreover, the signal progressively increases with $V_G$, while the position of the zero-signal along the NW axis remains practically unperturbed. Since $V_{DS} = 0$ V, we can only probe the photovoltage signal that is directly generated by photo-excitation, as in the case of the PTE or PW excitation, but we are not sensitive to any photo-induced variation of the conductivity, *i.e.*, to any bolometric effect.

Accordingly, we repeat the same measurements on NW2 as sketched in Fig. 3b while applying $V_{DS} = 1$ mV at the S contact, corresponding to a *DC* dark current of ~5 nA. The near-field maps and line profiles, corresponding to the second-order ($n = 2$) component of the photocurrent $\Delta I_2$, are shown in Fig. 3d and 3f, respectively. Since the signal is acquired from D, a positive photocurrent corresponds to a current flowing from S to D. At $V_G = 0$ V, as we illuminate NW2 with THz light, we observe a $\Delta I_2$ signal with constant polarity along the whole NW, with only small intensity modulations, which resembles the expected response for the bolometric effect plotted in Fig. 2d. Since the number of carriers contributing to the conduction is not affected by THz-light absorption, the conductance change associated with the observed bolometric photocurrent can be ascribed to a temperature-induced mobility variation. Interestingly, by increasing $V_G$ above the *pinch-off* ($V_G > 2.5$ V), the photocurrent maps drastically change, and the phase jump, distinctive of the PTE, reappears. The enhancement of the PTE, which eventually dominates at high carrier densities, overwhelming the bolometric response, is encoded in the dependence of $S_e$ on the carrier density $n_e$, which determines the increase with $V_G$. To elucidate the interplay between the bolometric effect and the PTE, we explore the dependence of the photocurrent signal in NW2 in two opposite regimes: when the photo-response is dominated by the PTE ($V_G = 6$ V) and when the bolometric effect is dominant ($V_G = 0$ V). Fig. 4 shows the $\Delta I_2$ maps and the related horizontal cuts as a function of the incident THz-field power $P$ and as a function of $V_{DS}$. The incident power is progressively increased by varying the bias applied to the THz-QCL heterostructure, and it is measured with a calibrated power meter (Thomas Keating THz Absolute Power & Energy Meter System) placed at the entrance of the s-SNOM instrument. To better quantify how the incident power affects the retrieved near-field signals, we integrated the horizontal line profiles, extrapolating the area underneath. In the PTE regime (Fig. 4 a-f), the signal increases with the incident power (see Fig. 4e) and remains almost constant while varying $V_{DS}$ up to $V_{DS} = 10$ mV (see Fig. 4f). Conversely, in the bolometric regime, see Fig. 4g-4l, the near-field photocurrent increases linearly both with the incident THz power (see Fig. 4k) and with $V_{DS}$ (see Fig. 4l).

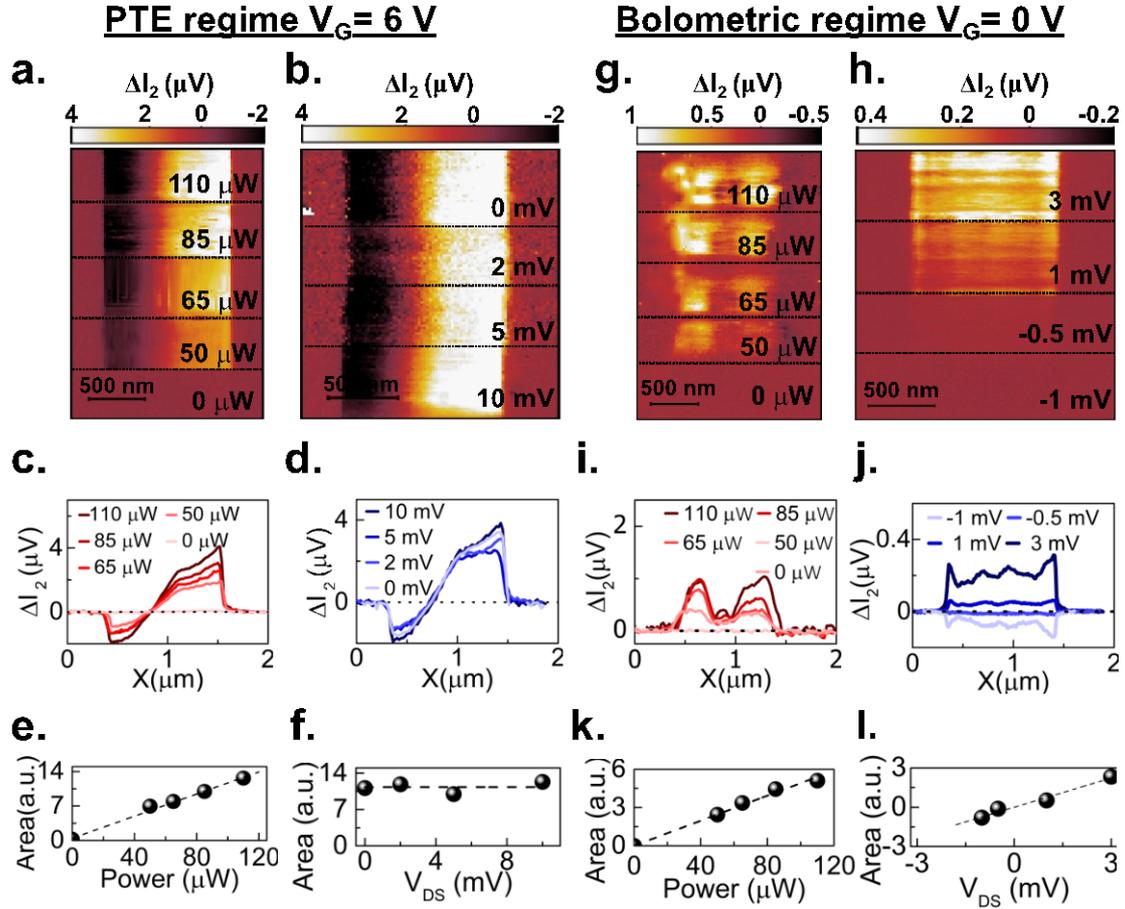

**Figure 4**: Evolution of the retrieved photocurrent signals. **(a-b)** Second-order photocurrent $\Delta I_2$ maps of NW2 in the PTE regime (gate voltage $V_G$ = 6 V) as a function of the incident THz power $P$ at fixed source-drain bias $V_{DS}$ = 5 mV (a) and as a function of $V_{DS}$ at a fixed $P$ of 100 μW (b). **(c-d)** $\Delta I_2$ linear profiles in the PTE regime extracted by averaging horizontal cuts of panels (a-b) in a 50 nm range around the NW centre. **(e-h)** Integrated areas in the PTE regime underneath the $\Delta I_2$ profiles shown in panels 4c-4d. **(g-h)** $\Delta I_2$ maps of NW2 in the bolometric regime (gate voltage $V_G$= 0 V) as a function of $P$ at a fixed $V_{DS}$ of 10 mV (g) and as a function of $V_{DS}$ at a fixed $P$ of 100 μW (h). **(i-j)** $\Delta I_2$ linear profiles in the bolometric regime, extracted by averaging horizontal cuts of panels (g-h) in a 50 nm range around the NW centre. **(k-l)** Integrated areas in the bolometric regime underneath the $\Delta I_2$ profiles shown in panels 4i-4j.

In addition to thermally driven effects, PW photo-excitation could generate $\Delta V$ and $\Delta I$ signals via the Dyakonov-Shur rectification[51] due to the capacitive coupling between the S and D contacts and the AFM tip. Since this coupling depends on the relative distance between the tip and the electrodes, it is influenced by tip tapping. Accordingly, the PW photo-response can, in principle, contribute to the demodulated signal at $n = 2$. The signal profile and photocurrent versus $V_G$ expected for the PW excitation is the same as that for the PTE[67], meaning that a phase change is expected at the NW centre. However, the PW photo-response is expected to increase with the applied $V_{DS}$ because of the forced injection of electrons in the FET channel and the enhanced asymmetry induced by the $I_{DS}$ current[11,68]. Accordingly, the lack of dependence on $V_{DS}$ in the regime where we observe the phase change (Fig. 4d) clearly indicates that the PW is here negligible and overwhelmed by the PTE response.

In all the explored conditions, the PTE photocurrent is at least four times more intense than the bolometric photocurrent; however, due to the different dependences on $V_{DS}$, the bolometric contribution can eventually become comparable to the PTE current even at high carrier density ($V_G$ above the pinch-off) if $V_{DS}$ is sufficiently high. This is clearly probed by measuring the $V_G$-dependent photocurrent maps at $V_{DS} = 10$ mV (see Supplementary Information). At gate voltages $V_G = 4$ V and $V_{DS} = 10$ mV, the two physical effects (photo-thermoelectric and bolometric) become comparable, leading to a higher detected total photocurrent at the expense of a mixed detection regime. The dominant role of the PTE is not surprising considering the giant thermoelectric response demonstrated in InAs NWs[66].

Detectors relying on the PTE can operate without the application of any current; conversely, for bolometers, the detector current represents an additional source of noise. To take advantage of the strong PTE response of InAs NWs, the coupling elements used to funnel the THz light into the nanoscale active region can be designed to favour the formation of thermal gradients through asymmetric photoexcitation of the active element. This is achieved by using a nanoscale antenna with one arm coupled to the gate and the other to the source or drain electrode.

The carrier density has a paramount role in defining the mechanism that dominates the photocurrent generation in InAs NWs at THz frequencies and can be quantitatively determined from nano-optical property measurements[13,35,69].

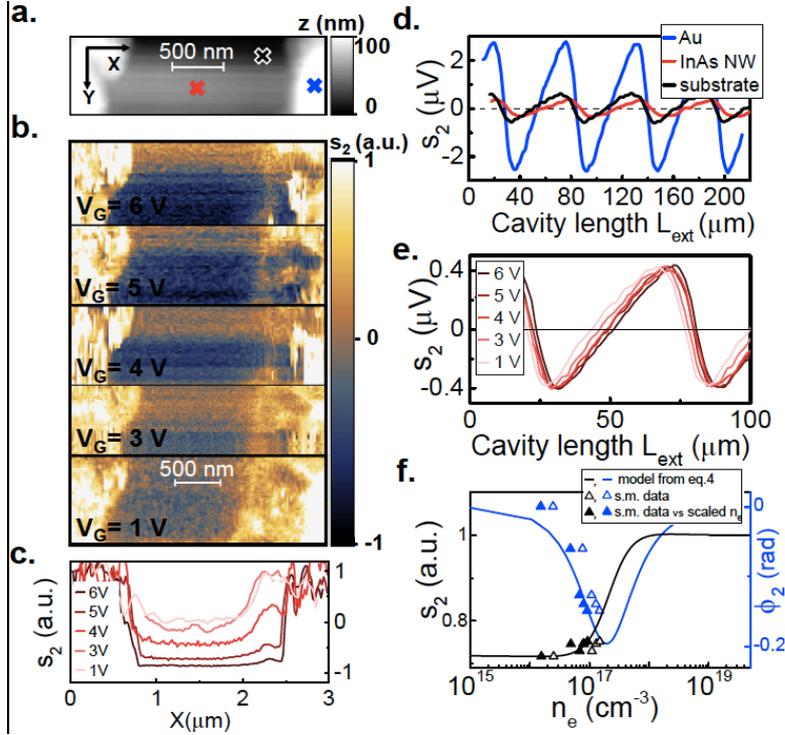

**Figure 5:** Detectorless near-field optical microscopy. (a) Topography of the NW1 nanowire. (b) Self-mixing intensity maps demodulated at $2\Omega$ ($s_2$) at different $V_G$ values scaled to the substrate value and normalized to the value measured at the gold contacts. The maps are acquired at fixed $L_{ext}$ =33.5 µm, corresponding to the maximum of the individual self-mixing fringes on gold electrodes. (c) $s_2$ line profiles obtained by integrating horizontal cuts of the $s_2$ maps in a 50 nm range around the NW1 centre and at different $V_G$ values. (d) Self-mixing fringes as a function of the external cavity length $L_{ext}$ at $V_G$ = 0 V. Individual curves correspond to different positions on the sample, marked as coloured crosses in panel (a): blue for the gold electrode, red for NW1, black for the substrate. (e) Self-mixing fringes acquired at a fixed position on NW1 as a function of the external cavity length $L_{ext}$ for different gate voltages $V_G$. (f) (f) Amplitude $s_2$ (black) and phase $\varphi_2$ (blue) of the second-order demodulation self-mixing signal of panel (e) as a function of the carrier density $n_e$ estimated from the transconductance curve assuming a constant mobility (open triangles), together with the predictions of eq. (4) (solid lines). The experimental $s_2$ values are scaled by a factor of 0.55 µV$^{-1}$ to match the model at $V_G$ = 1 V. Better agreement between the experiment and model is observed for carrier densities $n_e$ scaled by a factor of 1.6 (full triangles).

To this aim, we exploit the capability of THz-QCL to serve both as a powerful source and as a phase-sensitive detector to monitor the change in field scattered by the InAs NW as a function of the applied $V_G$. The THz light back-scattered by the AFM tip eventually returns into the laser cavity, perturbing its emission via the self-mixing effect[61,70]. We isolate the near-field contribution to the back-scattered light by monitoring the QCL contact voltage $V_{QCL}$ and demodulating it at $\Omega_n$. The $s_n$ signal from InAs NWs has generally lower intensity than those captured from the undoped Si substrate and from the gold electrodes, as seen in the map of the second-order component $s_2$ in Fig. 5b and the corresponding topography (Fig. 5a), where the NW appears as a dark stripe bridging the much brighter gold contacts. The self-mixing signal $s_2$ is a function of the length $L_{ext}$ of the external cavity defined by the AFM tip and by the laser emission facet such that self-mixing fringes can be acquired by changing $L_{ext}$ with a delay line in the optical path. The fringes collected at three different spots on the sample corresponding to the NW, substrate and electrodes are shown in Fig. 5d. The phase of the back-scattered field ($\varphi_n$) is directly related to the phase of the self-mixing fringes.

By scanning the sample, we observe a change in the amplitude and degree of asymmetry of the self-mixing fringes due to the variation in the scattered intensity, which is fed back into the laser cavity. The self-mixing contrast is evaluated by subtracting the signal from the substrate and by normalizing the maps to the value retrieved on the gold electrodes. The line profiles are extracted by averaging the horizontal cuts of the maps of Fig. 5b in a 50 nm range centred on the NW. Negligible variations are seen by changing the current flowing in the NW with $V_{DS}$ (see Supporting Information). In contrast, as $n_e$ increases with $V_G$, we observe a clear change in the near-field contrast at a fixed $L_{ext}$ (see Figs. 5b-5c). This change can be attributed to variation in the dielectric permittivity $\varepsilon$ of the NW with carrier density $n_e$, as described by the Drude model.

The field scattered by the tip is indeed related to the incident field by the complex-valued scattering coefficient $\sigma_s = (1+r_p)^2 \alpha_{eff}$, where $r_p$ is the Fresnel coefficient accounting for the reflection by the sample surface and $\alpha_{eff}$ is the effective polarizability that depends on the sample-tip near-field interaction and is the function $\varepsilon$. We can expand the second-order scattered signal $s_2$ in $\varepsilon$ in the framework of the point dipole moment model,[71] as detailed in the Methods section, to trace the changes in $s_2$ to the dependence of $\varepsilon$ on $n_e$.

In the investigated carrier density range, the model predicts a small variation in the signal amplitude and a consistent shift of the phase of the scattered field. To disentangle the effects of carrier injection on the phase ($\varphi_2$) and amplitude ($s_2$) of the scattered field, we acquire the self-mixing signal of NW1 as a function of $L_{ext}$ at different gate voltages, as reported in Fig. 5e. While $s_2$ remains barely unchanged at $V_G$ values up to 6 V, a phase drift is observed, a signature of a signal depletion at increasing $V_G$ in the maps, at constant $L_{ext}$ in Fig. 5b. The phase shift can be attributed to the photoexcitation of plasmon resonance of the NW[13], whose frequency matches the probe photon energy at carrier densities on the order of $10^{17}$ cm$^{-3}$. To compare the observed trend with predictions based on the Drude model of $\varepsilon$ of the InAs NW (see Methods), we evaluate the carrier density $n_e$ corresponding to the applied $V_G$ from the transconductance curve in the range $V_G$ = 1-6 V while assuming constant mobility (Fig. 5f). The signal amplitude s2 is normalized to the predicted value for $V_G$ = 1 V to rule out any dependence on the acquisition parameters (alignment, self-mixing parameters, THz coupling to the tip). The model predicts the observed trends of signal amplitude and phase, but it indicates that $n_e$ is overestimated. A better agreement is reached by dividing the extracted $n_e$ values by a factor of 1.6, meaning that the carrier mobility is slightly underestimated here (see Fig. 5f).

The carrier density $n_e$ corresponding to the applied $V_G$ is determined from the transconductance curve in the range $V_G$ = 1-6 V, assuming constant mobility. The more carriers that are injected in the NW, the greater the plasmon resonance of the NW[13] approaches the probe photon energy, resulting in a phase shift of the field scattered back into the QCL to produce the self-mixing signal. To rule out any dependence of the absolute value of $s_2$ on the acquisition parameters (alignment, self-mixing parameters, THz coupling to the tip), the signal reported in Fig. 5f is normalized to the predicted value for $V_G$ = 1 V (green open circle).

The THz response at 2.7 THz shows a strong dependence on the carrier density $n_e$ [34] in the range $10^{16}$ - $10^{18}$ cm$^{-3}$, and the trend of the self-mixing signal of the NW indicates that at $V_G$ up to 6 V, the gating effect is not saturated, as it can occur when the charge carriers in the NW shell screen the gate potential, and we are then still able to inject carriers into the NW. Finally, by exploiting the imaging capability of s-SNOM, we observe that the self-mixing signal in Fig. 5b,c varies along the NW axis, exhibiting a minimum contrast near the D contact, which could be induced by local variations in the carrier density caused by crystal impurities.

**Discussion**

In conclusion, we exploit photocurrent nanoscopy at THz frequencies to identify the photo-detection mechanisms in InAs NW FETs. Once the radiation of a THz QCL source impinges on the s-SNOM AFM tip, we capture snapshots of the photocurrent flowing in the strongly subwavelength NW section, and we retrieve clear signatures of two thermally driven processes: the PTE and the bolometric effect. The interplay between these two mechanisms is discussed as a function of the carrier density, $V_{DS}$ and incident THz power.

Interestingly, at $V_{DS} = 0$ V, the PTE dominates at high carrier density. Such a condition is extremely advantageous in engineering RT photodetectors combining low noise equivalent powers and high speeds. Indeed, the zero-bias operation reduces dominant noise sources such as flicker noise, shot noise, and generation-recombination noise, owing to the absence of a static current. Moreover, the high carrier density operation entails a reduction of the channel resistance (the NW can be operated at its maximum conductivity state), which, in turn, results in a proportional reduction in the overall RC time constant. Eventually, the positive correlation between the device temperature and the Seebeck coefficient, given by the Mott relation, makes the PTE mechanism ideal for RT operation. On the other hand, to take full advantage of the bolometric detection mechanism, low temperatures are preferable because of the reduced lattice thermal conductivity and suitable device architectures capable of minimizing the thermal exchange between the NW and its surroundings, such as suspended NW architectures. The achieved results provide a route for engineering THz RT photodetectors with larger noise equivalent powers (NEPs) and inherently high speeds, which could be attained by favouring the formation of light-induced thermal gradients along the NW axis to exploit the highly efficient photo-thermoelectric response of InAs NWs and by the inherently small (attofarad) capacitance. In InAs NWs, knowledge of the local photocurrent can be used to select the ideal position of the lateral gate in a planar NW geometry, where the gate is coupled to one of the arms of a dipole antenna.[14] In turn, this choice can simultaneously maximize the photocurrent and determine the dominant detection mechanism. By maximizing the photocurrent, the noise equivalent power can be reduced. [21,72]

On the other hand, it is well known that a specific detection mechanism directly affects both the NEP and the response time of a detector, and this is valid for different material platforms. [73-75] Indeed, a device operating via the PTE, not requiring a direct bias through the channel (unlike what happens in a bolometer[73,75,65]), is less prone to a number of noise sources; flicker noise, generation-recombination noise and shot noise all scale with increasing current, eventually dominating the noise figure of a bolometer even at RT. Therefore, THz photocurrent nanoscopy measurements indicate a way to reduce NEP by identifying a route to selectively activate the PTE.

Selective activation of the PTE detection mechanism would also allow increasing the detection speed, which, in a thermally driven THz photo-response, is limited by the efficiency of the energy transfer between incoming photons and the photo-detecting system. In the case of the PTE, the photo-detecting system is the electronic thermal distribution that is modified by the absorption of a photon.[65] The characteristic timescale of the photo-response is then given by the electronic specific heat, which limits the rise-time of the signal, and by the electron cooling rate (mainly towards the phonon bath). On the other hand, the bolometric effect is driven by a change in lattice temperature, which occurs subsequent to the formation of a temperature gradient in the electronic distribution,[65] i.e., the lattice temperature changes as a consequence of the increase in the electronic temperature, in agreement with the model described in Fig. 2.

The proposed THz nanoscopy investigation will open up research opportunities in many application domains: real-time pulsed imaging and time-of-flight tomography; time-resolved THz spectroscopy of gases, complex molecules and cold samples; coherent control of quantum systems; quantum optics, where high-power pulses can drive molecular samples out of equilibrium and ultrafast detectors can capture such an effect; metrology; and ultra-high-speed communications, where THz frequency carriers will become increasingly more important in the quest for higher bandwidth data communications and finally can promise extraordinary impacts on the market for biomedical imaging, security and process control. The photocurrent distributions and polarities corresponding to the two effects are consistently different, making photocurrent nanoscopy a very valuable tool in investigating photo-conduction in low-dimensional materials.

Furthermore, we demonstrate that a detectorless s-SNOM system built with a THz-QCL can serve as a multimodal near-field microscope, providing complementary information on optical properties via self-mixing interferometry. The strong dependence of the THz scattered intensity makes this technique particularly useful to probe the Drude response to THz fields for charge carrier densities in the $10^{16}$-$10^{18}$ cm$^{-3}$ range.

**Materials and Methods**

**Sample fabrication**
Au-catalysed n-doped InAs NWs are grown by chemical beam epitaxy (CBE) on InAs (111) substrates using trimethylindium (TMIn) and tertiarybutylarsine (TBAs) and ditertiarybutylselenide (DtBSe) as metal-organic (MO) precursors. As-grown InAs NWs have an average length of 2.3±0.5 μm and radii in the 40-120 nm range. NW FETs are realized with standard nano-fabrication methods. NWs are transferred from the growth substrate to the host substrate *via* dry transfer. The host substrate is a 350 μm thick intrinsic silicon wafer capped by 500 nm SiO$_2$. Individual NWs are then geometrically analysed with SEM imaging, and electrical contacts are defined by aligned electron beam lithography (EBL). Before metal evaporation, a chemical wet passivation step is performed on the exposed NW areas. The ammonium polysulfide (NH$_4$)$_2$S$_x$ solution employed for this step removes the oxide layer that covers the NW before metallization, thus ensuring Ohmic metal-semiconductor contacts. The lateral-gated FETs are then finalized by the thermal evaporation of a 10/100 nm Cr/Au layer and *lift-off*.

**Near-field microscopy**

The THz-QCL is mounted in a liquid helium continuous-flow cryostat with a polymethylpentene window and maintained at a fixed heat sink temperature of 13 K. The emitted THz beam is collimated using a 90 degrees off-axis parabolic mirror with an effective focal length of 50 mm. The collimated THz beam is fed into the entrance optical port of a commercial near-field microscope (NeaSNOM, Neaspec, Martinsried, Germany). The optical path length is varied using a delay line, with a linear translation stage having a 30 nm precision. In the microscope, a second paraboloid mirror with an equivalent focal length of 25 mm focuses the beam onto a PtIr tip (RMN-25PtIr300B-H10) having a nominal apex radius of 10 nm and a shank length of 80 μm that is resonant at tapping frequency $\Omega$ = 20 (±30%) kHz. The laser polarization lies in the plane containing the tip to efficiently induce an oscillating dipole in the tip. The metallized AFM probe in close proximity to the NW can influence the potential in the transistor channel and thus photocurrent generation by screening the gate field. To minimize this impact, we do not electrically ground the metal cladding of the tip. To maximize the sensitivity to coherent optical feedback, in the self-mixing measurements, the QCL is driven in a continuous wave at the current I = 780 mA (J = 256 Acm$^{-2}$), which is just 5% higher than the threshold current density ($J_{th}$ = 240 Acm$^{-2}$), using a highly stable current generator (Lightwave Electronics, mod. QCL 2000). To perform photocurrent nanoscopy, the QCL is driven in a continuous wave with a current in the range I = 780 mA-830 mA. The THz radiation scattered by the sample is collected by the same focusing parabolic mirror and coupled back to the laser cavity along the same incident optical path. The voltage modulation across the QCL terminals produced by the self-mixing effect is pre-amplified using a low-noise amplifier (DL Instruments, mod. 1201) and demodulated up to the highest harmonic order (*n* = 5) allowed by the electronic card of the NeaSNOM. The average optical path length from the QCL front facet to the tip is 60 cm. Approach curves reveal that the second harmonics (*n* = 2) of the electrical signals are already free from far-field background contribution (see Supporting Information), and they are chosen as experimental result metric.

**Near-field self-mixing signal and carrier density estimation**

In the framework of the point dipole moment model,[63,71,76] the effective probe polarizability $\alpha_{eff}$ can be expressed in terms of the electrostatic coefficient $\alpha$, which is the dipole-polarizability, and $\beta$, which is the quasi-static reflection coefficient:

$$\alpha_{\textit{eff}} = \frac{\alpha}{1 - \dfrac{\alpha\beta}{16\pi(R+h)}} \quad (2)$$

where $h$ is the sample-probe distance and $R$ is the tip radius. To estimate the second-order scattering contribution to the s-SNOM signal $s_2$, we consider the second-order expansion of $\alpha_{\textit{eff}}$ at a power of $h$, which is varied with a sinusoidal function at a frequency equal to the tapping frequency. The second-order coefficient is:

$$s_2 \propto -\frac{3}{128}\alpha^2\beta(1+\beta)(8+\alpha\beta) \quad (3)$$

We can expand eq. (3) in the dielectric constant $\varepsilon$ of the NW considering that $\beta = \dfrac{(\varepsilon-1)}{\varepsilon+1}$ and that for a metal-coated tip $\alpha = \dfrac{(\varepsilon-1)}{\varepsilon+2} \approx 1$. The expression of the complex-valued $s_2$ simplifies as:

$$s_2 \propto -3\varepsilon\frac{(7+2\varepsilon-9\varepsilon^2)}{(3\varepsilon+5)^3} s_2 \quad (4)$$

The permittivity is described by the Drude model as $\varepsilon = \varepsilon_\infty\left(1 - \dfrac{\omega_{pl}^2}{\left(\omega^2 - \dfrac{i\omega}{\tau}\right)}\right)$, where the high-frequency dielectric constant $\varepsilon_\infty = 13.13$ as in Ref. 13; $\omega_{pl}$ is the plasma frequency, which depends on the carrier density $\omega_{pl} = \sqrt{\dfrac{4\pi n_e}{\varepsilon_\infty m^*}}$; $m^*$ is the carrier effective mass ($= 0.023\, m_e$); $m_e$ is the free electron mass; $\tau = 10$ fs is the Drude scattering time as in Ref. 13 and $n_e$ is the carrier density tuned with the lateral gate.

**Acknowledgements**

This work was supported by the European Research Council through the ERC Consolidator Grant (681379) SPRINT, by the European Union through the H2020-MSCA-ITN-2017, TeraApps (765426) grant, partially by the SUPERTOP project of the QuantERA ERA-NET Cofund in Quantum Technologies and by the FET-OPEN project. M.S.V. acknowledges partial support from the second half of the Balzan Prize 2016 in applied photonics delivered to Federico Capasso. The authors acknowledge fruitful discussions with A. Tomadin.

**Supporting Information Available**

The following files are available free of charge. SupportingInfo.pdf

**Conflict of interests**

Competing financial and non-financial interests: The authors declare no competing financial and non-financial interests.

**Materials & Correspondence**

Correspondence and requests for materials should be addressed to E.A.A.P. (eva.pogna@nano.cnr.it) and M.S.V. (miriam.vitiello@sns.it)